\documentclass[namedreferences]{solarphysics}
\usepackage[hyperref,optionalrh,solaromanenum]{spr-sola-addons} % For Solar Physics 
\usepackage{graphicx}                    % For eps figures, newer & more powerfull
\usepackage{epstopdf}
\usepackage{color}                       % For color text: \color command
\usepackage{breakurl}                         % For breaking URLs easily trough lines
\chardef\us=`\_

\begin{document}

\begin{article}

\begin{opening}

\title{Solar models with dynamic screening and early mass loss tested by helioseismic, astrophysical, and planetary constraints}

%%%%%%%%%%%%%%%%%%%%%%%%%%%%%%%%%%%%%%%%%%%%%%%%%%%
%% Authors Names
%
 \author[addressref={aff1},corref,email={suzannah@lanl.gov}]{\inits{S.R.}\fnm{Suzannah R. }\lnm{Wood}\orcid{0000-0002-7208-7681}}
 \author[addressref={aff1},corref,email={mussack@lanl.gov}]{\inits{K.}\fnm{Katie }\lnm{Mussack}\orcid{0000-0002-5539-9034}}
 \author[addressref={aff2},corref,email={joy@lanl.gov}]{\inits{J.A.}\fnm{Joyce A. }\lnm{Guzik}\orcid{0000-0003-1291-1533}}

%%%%%%%%%%%%%%%%%%%%%%%%%%%%%%%%%%%%%%%%%%%%%%%%%%%
%% Runningheads
%
\runningauthor{Wood, Mussack, \& Guzik}
\runningtitle{Dynamic Screening and Early Mass Loss}

%%%%%%%%%%%%%%%%%%%%%%%%%%%%%%%%%%%%%%%%%%%%%%%%%%%
%% Affilations 
%% id shold be the same with \author addressref value.
\address[id={aff1}]{Los Alamos National Laboratory, XTD-IDA, MS T-086, Los Alamos, NM  87545 USA}
\address[id={aff2}]{Los Alamos National Laboratory, XTD-NTA, MS T-082, Los Alamos, NM  87545 USA}

%%%%%%%%%%%%%%%%%%%%%%%%%%%%%%%%%%%%%%%%%%%%%%%%%%%
%%% Abstract 
\begin{abstract}
The faint young Sun paradox remains an open question. Presented here is one possible solution to this apparent inconsistency, a more massive early Sun. Based on conditions on early Earth and Mars, a luminosity constraint is set as a function of mass. We examine helioseismic constraints of these alternative mass-losing models. Additionally, we explore a dynamic electron screening correction in an effort to improve helioseismic agreement in the core of the an early mass-losing model.  
\end{abstract}

%%%%%%%%%%%%%%%%%%%%%%%%%%%%%%%%%%%%%%%%%%%%%%%%%%%
%% Keywords
%

\end{opening}
%-------------------------------------------------

%%%%%%%%%%%%%%%%%%%%%%%%%%%%%%%%%%%%%%%%%%%%%%%%%%%%%%%%%%%%%%%%%%%%%%%%%%%
%% Introduction
\section{Introduction}

There is evidence for the presence of liquid water on Earth as early as 4.3 Gyrs ago, and on Mars as early 3.8 Gyrs ago. However, for 1 $M_{\odot}$ standard solar evolution modes, the solar luminosity was 70\% of the current solar luminosity at the beginning of the Sun's main-sequence lifetime, which is not enough for the presence of liquid water at these early times. This inconsistency is known as the ``faint young Sun paradox." There are many proposed ways to account for this apparent contradiction (\textit{e.g.}, greenhouse gases effects \citep{Forget_2013, Turbet17, Bristow17, Wordsworth16}, carbon cycle \citep{Charnay17}, coronal mass ejections \citep{Airapetian16a}, more massive early Sun \citep{SB03, Minton2007, STC11, WH13}). 

It is possible that higher mass-loss rates seen in pre-main sequence stars extend into the early main sequence, and reasonable that the mass-loss rate decreases rather quickly with time on the main sequence \citep[\textit{e.g.}, power law or exponentially as used by \textit{e.g.},][]{Wood02, Wood05a, Wood05b, Ribas10, GWB87}. Therefore, the Sun could have begun its main-sequence evolution at higher mass and luminosity \citep[as investigated by \textit{e.g.},][]{GWB87, STC88, SB03, Minton2007, STC11, WH13}.

Here, we use the solar luminosity requirement suggested by the presence of liquid water on early Earth and Mars to place limits on the initial mass of a more massive, luminous Sun, assuming an exponentially decaying mass-loss rate. We examine the interior sound speed and oscillation frequencies of these alternate solar models and compare with helioseismic constraints.  

As found by \citet{GM2010}, early mass loss can improve sound-speed agreement with helioseismic inferences in the radiative interior outside the core H-burning region, but worsens the agreement in the solar core. \citet{Mussack_2011} found that taking into account electron screening corrections to nuclear reaction rates based on molecular dynamics simulations resulted in a change in core sound speed in the opposite direction to the effects of early mass loss. Here we investigate the compensating effects on core structure of including both early mass-loss and dynamical electron screening corrections simultaneously. 

%%%%%%%%%%%%%%%%%%%%%%%%%%%%%%%%%%%%%%%%%%%%%%%%%%%%%%%%%%%%%%%%%%%%%%%%%%%
%% Solar Evolution Models
\section{Solar Evolution Models}
\label{sect: models}

The Los Alamos group models use a one-dimensional evolution code updated from the \citet{Iben_1963, Iben_1965a, Iben_1965b} code. Updates to the \citet{Iben_1963, Iben_1965a, Iben_1965b} code, are the same changes are implemented in \citet{GM2010}. These incluse use of the OPAL opacities \citep{IR96} supplemented by the \citet{Fer05} or Alexander \& Ferguson (private communication, 1995) low-temperature opacities, SIREFF equation of state \citep[see][]{GS97}, Burgers' diffusion treatment \citep[for implementation see][]{CGK89}, and nuclear reaction rates from NACRE \citep{Angulo_1999}, with a correction to the $^{14}$N rate from \citet{Formicola_2004}. In this work the abundances of \citet{AGS05} and \citet{GN93} will be employed to bracket the range of abundance determinations.

For a full description of the physics, data, and codes used in the models discussed here, refer to \ \citet{GM2010} and \citet{GWC05}. For a description of the mass-loss function, refer to \citet{GWB87}.

Models were computed from the pre-main sequence contraction phase, when the luminosity generated from gravitational contraction exceeds the luminosity generated by nuclear reactions by a factor of several, through the present day. For mass loss models, the model was evolved at a constant mass until the zero-age main sequence, defined as a minimum in radius, at which point the mass loss was turned on. The evolution models were adjusted to present standard solar values within uncertainties: present solar age \citep[4.54$\pm$0.04 Gyr,][]{Guenther_1992}, radius \citep[6.9599$\times10^{10}$ cm,][]{Allen_1973}, mass \citep[1.989$\times 10^{33}$ g,][]{CT86}, luminosity \citep[3.846$\times 10^{33}$ erg/g,][]{Willson_1986}, and specified photospheric ratio (Z/X). This was done by modifying the initial heavy element mass fraction (Z), initial helium mass fraction (Y), and the mixing length to pressure-scale factor ($\alpha$); see Table \ref{Tab:ModelSum} for initial and final values from our models.

The mass loss function decays exponentially with an e-folding time of 0.45 Gyr. This simple mass-loss treatment is smooth, decreases with time, and most of the mass is lost at early times. This mass-loss treatment was chosen because low-mass stars (like solar-mass stars) lose angular momentum quickly during main sequence evolution \citep{Wolff97} via magnetized stellar winds \citep{Schatzman62, Weber67}, due to a lower specific angular momentum. This mass loss occurs predominantly in the early lifetime of the star at ages less than 1 Gyr, for solar-like stars \citep{GB13}. The present solar mass-loss rate does not affect the Sun's evolution \citep[2x10$^{-14}$ $M_{\odot}$][]{Feldman_1977}.  

The dynamic electron screening correction was implemented following the treatment of \citet{Mussack_2011}.
%%%%%%%%%%%%%%%%%%%%%%%%%%%%%%%%%%%%%%%%%%%%%%%%%%%%%%%%%%%%%%%%%%%%%%%%%%%%%
\begin{table}
\caption{Initial Conditions (mass, abundances), Mixing-length Parameter, and Final Conditions (Abundances, CZ Base)}
\label{Tab:ModelSum}
\begin{tabular}{rccccccc}     % define the column alignment
                           % l: left, c: center, r: right
  \hline                   % horizontal line
  \hline 
Model &$M_{o}$/$M_{\odot}$ & $Y_{o}$ & $Z_{o}$ & $\alpha$ & $Z/X$ & $Y_{CZ}$\tabnote{Seismically inferred values from Basu and Antia (2004): $Y_{CZ}$ = 0.2485 $\pm$ 0.0035} & R$_{CZB}$/$R_{\odot}$\tabnote{Seismically inferred values from Basu and Antia (2004): $R_{CZB}$/$R_{\odot}$ = 0.713 $\pm$ 0.001} \\
\hline
GN93 & 1.00 & 0.2693 & 0.0196 & 2.0324 & 0.0239 & 0.2414 & 0.7185 \\
AGS05 & 1.00 & 0.2570 & 0.0135 & 1.9832 & 0.0163 & 0.2272 & 0.7307 \\
ML107 & 1.07 & 0.2545 & 0.0135 & 1.9912 & 0.0169 & 0.2393 & 0.7295 \\
ML115 & 1.15 & 0.2528 & 0.0135 & 2.0104 & 0.0172 & 0.2348 & 0.7258 \\
ML130 & 1.30 & 0.2466 & 0.0135 & 2.0571 & 0.0178 & 0.2388 & 0.7217 \\
AGSpp0 & 1.00 & 0.2576 & 0.0135 & 2.0139  & 0.0158 & 0.2277 & 0.7296 \\
ML107pp0 & 1.07 & 0.2550 & 0.0135 & 1.9710 & 0.169 & 0.2359 & 0.7334 \\
ML130rx0 & 1.30 & 0.2476 & 0.0135 & 2.0302 & 0.0167 & 0.2291 & 0.7257 \\
  \hline
\end{tabular}
\end{table}

%%%%%%%%%%%%%%%%%%%%%%%%%%%%%%%%%%%%%%%%%%%%%%%%%%%%%%%%%%%%%%%%%%%%%%%%%%%
%% Results and Discussion
\section{Results and Discussion}
\subsection{Determining Luminosity Constraints Based on Planetary Conditions}
By examining the conditions on the planets at early times, useful constraints can be placed on the early-time luminosity of the Sun. Conditions on the early Earth can be used as a upper constraint on luminosity, while those on early Mars can be used as lower constraint on luminosity based on their relative positions to the Sun. 
According to the standard solar model, the Earth should have been too cold for liquid water for the entire Archaean period, approximately the first half of its existence \citep{Kasting_2010}; however, it was not. 

\begin{figure}    %%%%%%%%%%%%%%%%%% FIGURE 1
                                % includes the two top panels 
   \centerline{
               \includegraphics[width=0.5\textwidth,clip=]{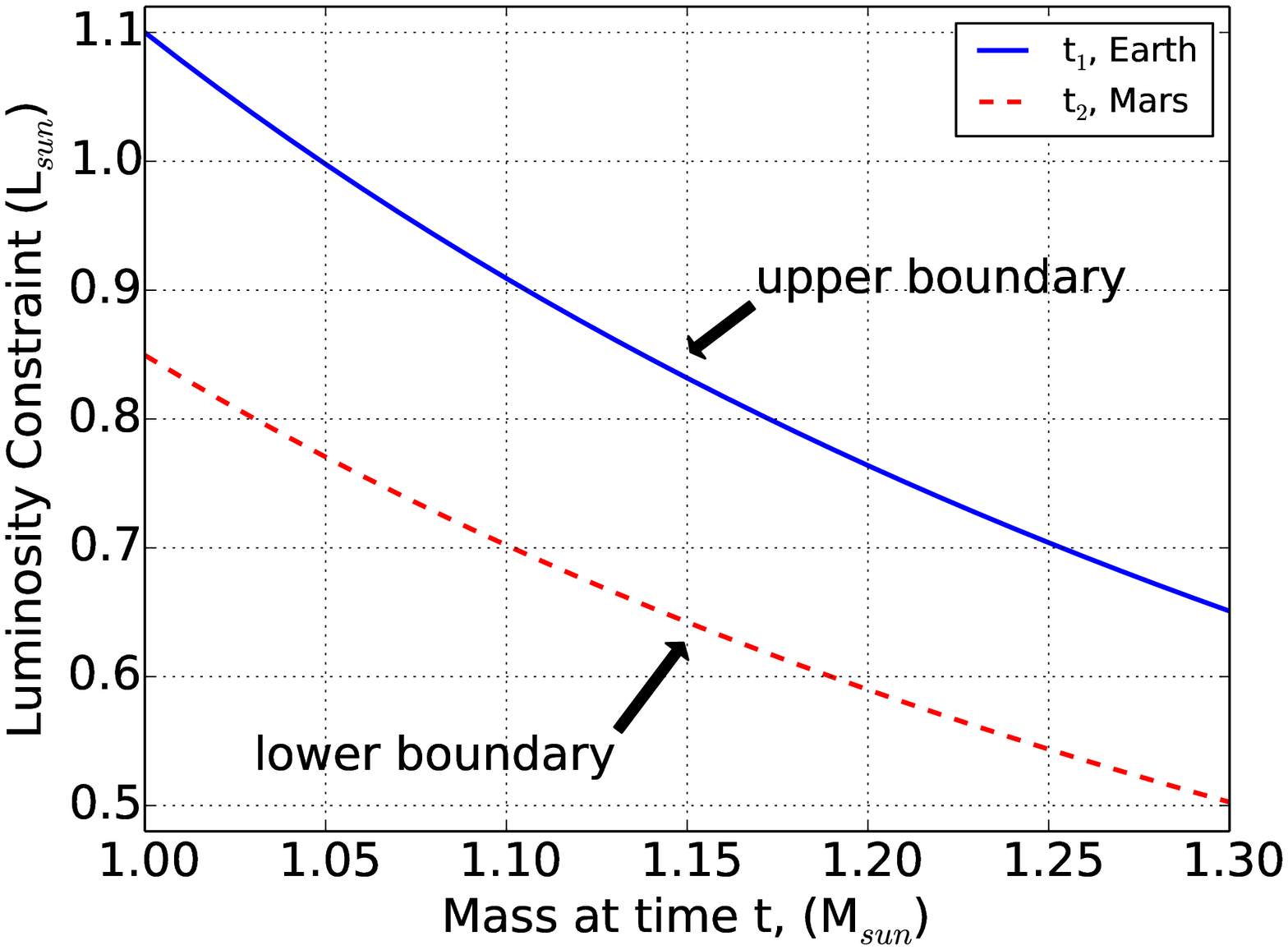}
               \includegraphics[width=0.5\textwidth,clip=]{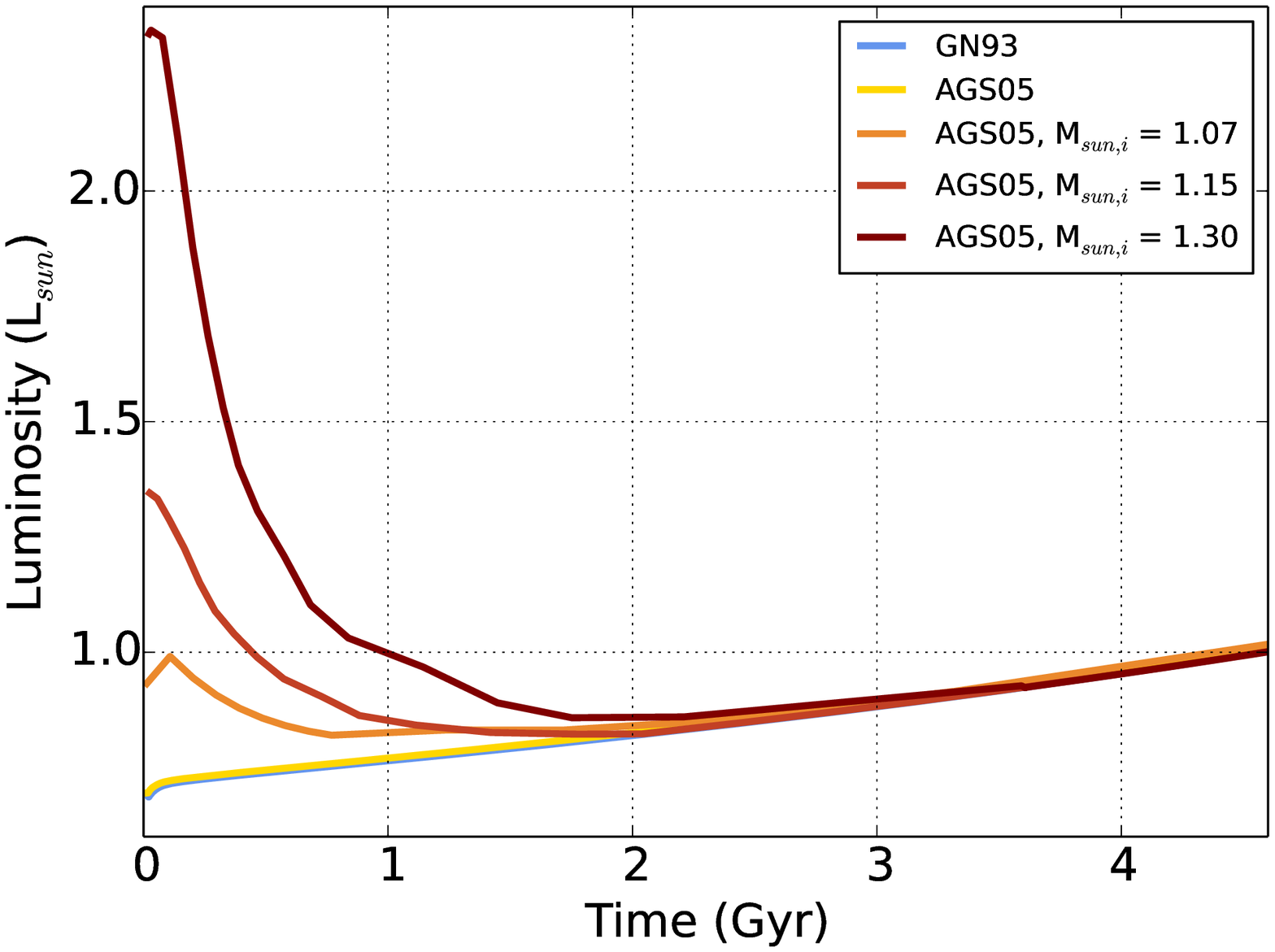}
              }
    \vspace{-0.405\textwidth}   % Shift close to the panel top 
     \centerline{ \bf     % Includes the labels (here needs the color 
                                %   package, see beginning of this file)
      \hspace{0.0\textwidth}  \color{black}{(a)}
      \hspace{0.415\textwidth}  \color{black}{(b)}
         \hfill}
      \vspace{0.35\textwidth}    % Shift back to the panel bottom

\caption{(a) Planetary constraints on luminosity as a function of solar mass at time (\textit{t}). Photodissociation of water from Earth at $t_{1}$ = 0.24 Gyr is \textit{blue}. Freezing of water on Mars at $t_{2}$ = 0.74 Gyr is \textit{red}. (b) Luminosity \textit{versus} time for solar models. Models include: standard solar models with GN93 (\textit{blue}) and AGS05 (\textit{yellow}) abundances, and early mass-losing models using the AGS05 abundances and initial masses of 1.07, 1.15 and 1.30 $M_{\odot}$ (\textit{orange to maroon} with increasing initial mass).
        }
   \label{fig:LvM}
   \end{figure}

The first evidence of a hydrosphere interacting with the Earth's crust was from 4,300 million years ago ($t$ = 0.24 Gyrs) \citep{MHP2001}. This suggests that the luminosity of the Sun needed to be low enough at this time to prevent the photodissociation of water and subsequent loss of H$_{2}$ into space. \citet{Whitmire_1995} provides constraints on mass loss based on the flux that reaches Earth at a given solar age. According to \citet{Kasting_1988} this would occur if the flux experienced by early Earth exceeds 1.1 $F_{\odot}$. If the average distance between the Earth and the Sun is assumed to be the same today as it was early in the Sun's evolution, then the Sun's maximum luminosity at early times could not exceed 1.1 $L_{\odot}$. However, if we were to consider a more massive Sun to explain the ``faint young Sun paradox," the distance between the early Earth and the early Sun would change as a function of the solar mass; therefore the luminosity to maintain liquid water on Earth would change as a function of mass. Even though the Earth's eccentricity varies by up to 6\% during the 100,000 year Milankovitch cycle, we assumed that the eccentricity of the Earth's orbit and the angular momentum per unit mass of the Earth remain constant with time, the average distance between the Earth and the Sun would be inversely proportional to the $M_{\odot}(t)$ at time $t$. Based on the condition that the flux experienced by early Earth does not exceed 1.1 $F_{\odot}$, Figure \ref{fig:LvM}a shows how the corresponding constraint on the luminosity varies with solar mass. The \textit{blue} line shows the upper limit on the solar luminosity at $t_{1}$ = 0.24 Gyr.

In addition, there is substantial geomorphological and geochemical evidence for liquid water on Mars at the end of the late bombardment period \citep{Carr_1996, ME00, ME03, Ehlmann_2008, Boynton_2009, Morris_2010, Osterloo_2008, Lanza16, Poulet_2005, Bibring_2006, Ehlmann_2011,Gendrin_2005, Squyres_2004, Squyres_2008, Bristow17, Turbet17}. In fact, \citet{Kasting_1991} and \citet{Kasting_1993} suggested that the solar flux needed to be 13\% greater than predicted by the standard solar model at the end of the late heavy bombardment period (3.8 Gyr ago, $t_{2}$ = 0.74 Gyr) in order for Mars to be warm enough for liquid water. However, if we were to consider a more massive Sun to explain the ``faint young Sun paradox," the distance between early Mars and the early Sun would also change as a function of solar mass; thus the luminosity to maintain liquid water on Mars would change as a function of solar mass. Assuming the eccentricity of the orbit of Mars and the angular momentum per unit mass of Mars remain constant with time, this lower boundary on luminosity is the red line in Figure \ref{fig:LvM}a. 

\begin{figure}    %%%%%%%%%%%%%%%%%% FIGURE 2
                                % includes the two top panels 
   \centerline{
               \includegraphics[width=0.5\textwidth,clip=]{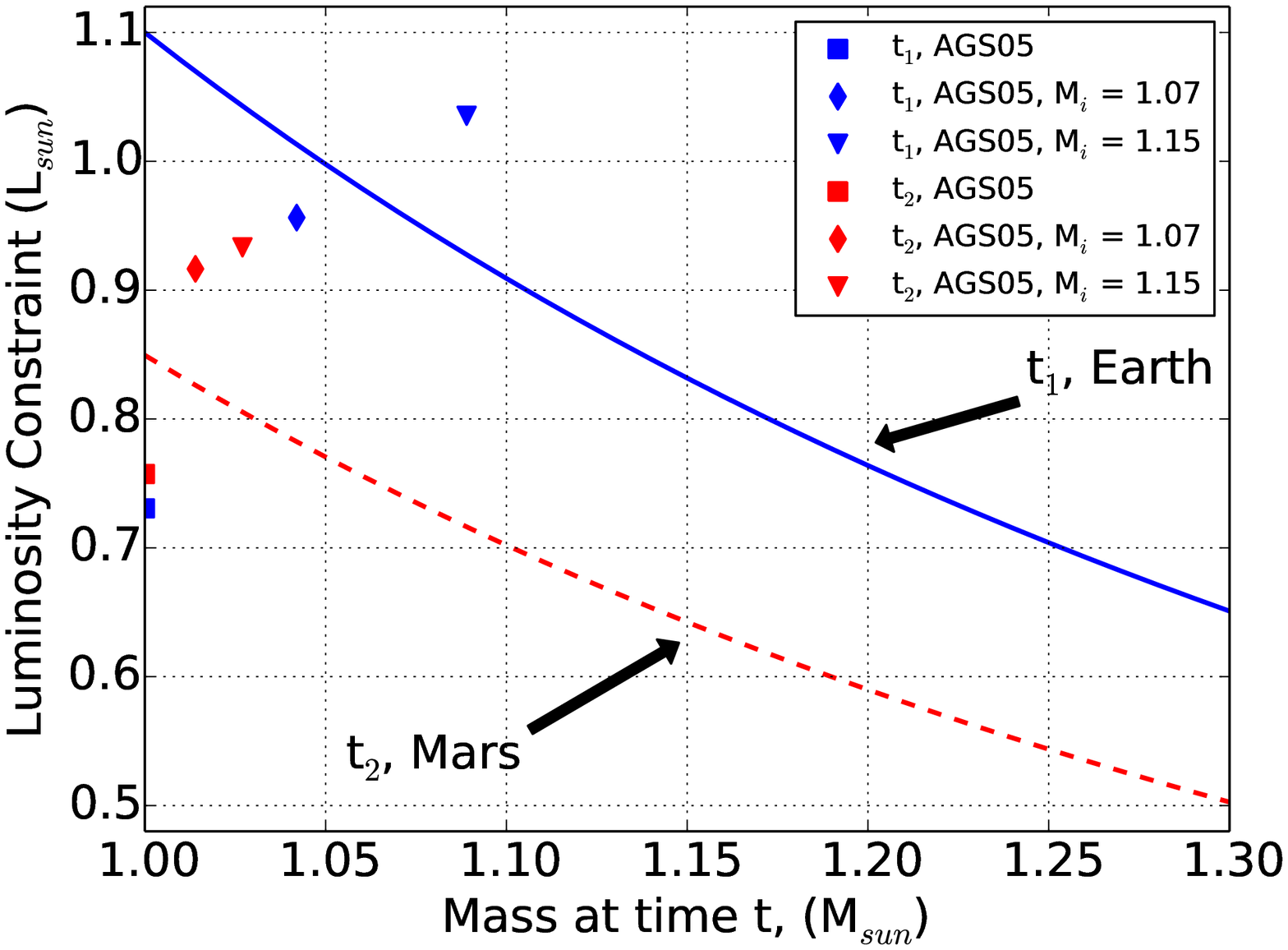}
               \includegraphics[width=0.5\textwidth,clip=]{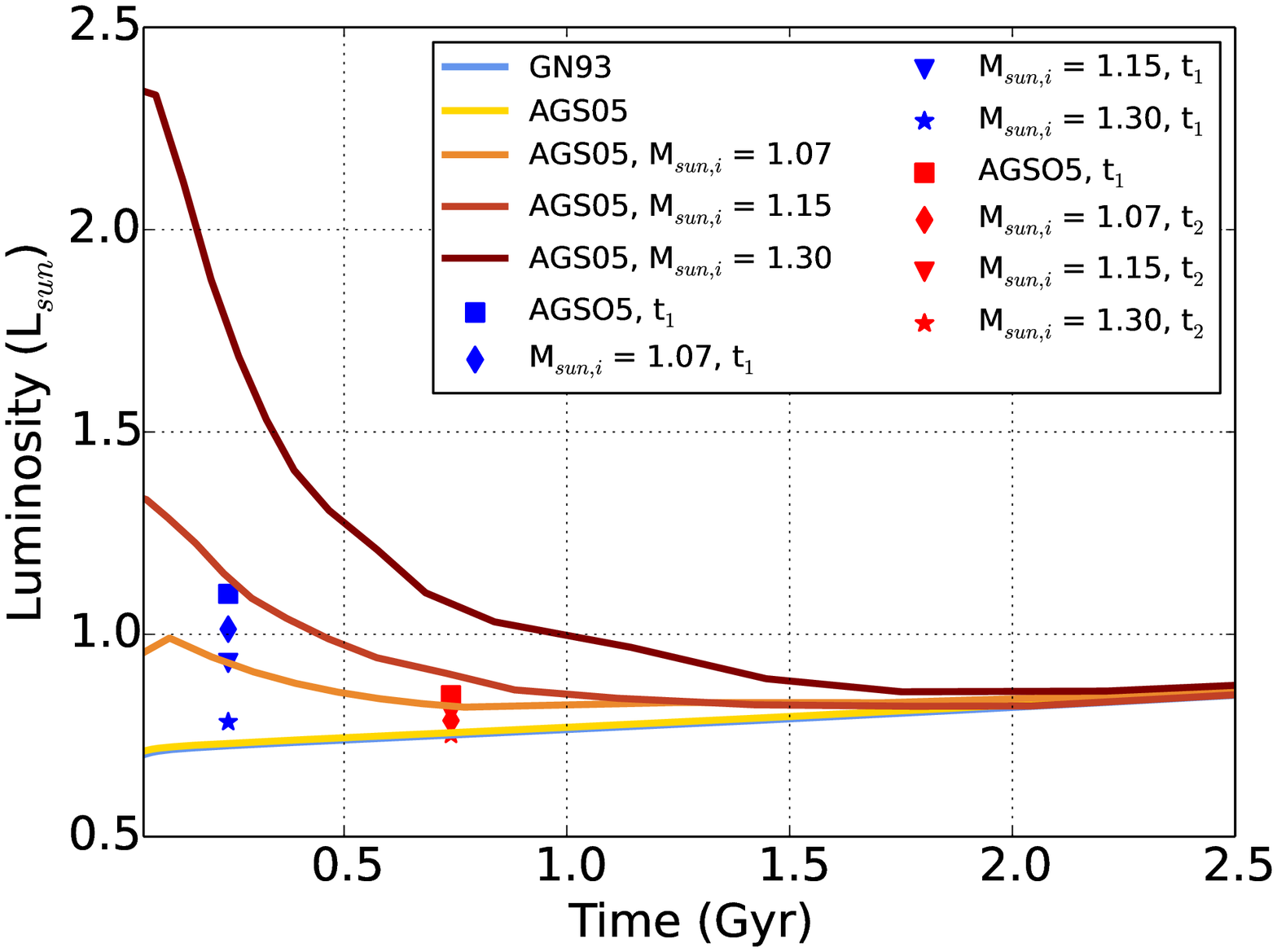}
              }
    \vspace{-0.405\textwidth}   % Shift close to the panel top 
     \centerline{ \bf     % Includes the labels (here needs the color 
                                %   package, see beginning of this file)
      \hspace{0.0\textwidth}  \color{black}{(a)}
      \hspace{0.415\textwidth}  \color{black}{(b)}
         \hfill}
      \vspace{0.35\textwidth}    % Shift back to the panel bottom

\caption{(a) Planetary constraints on luminosity as a function of solar mass at time (\textit{t}). The values of mass and luminosity are plotted for the models with initial mass of 1.00 (\textit{square}), 1.07  (\textit{diamond}) and 1.15 (\textit{triangle}) $M_{\odot}$ at  $t$ = $t_{1}$ (\textit{blue}) and $t$ = $t_{2}$ (\textit{red}). (b) Zoom-in of luminosity \textit{versus} time. With constraints on solar luminosity for the standard solar model (\textit{square}), and mass-losing models with starting masses of 1.07 $M_{\odot}$ (\textit{diamond}), 1.15 $M_{\odot}$ (\textit{triangle}), and 1.30 $M_{\odot}$ (\textit{star}).
        }
   \label{fig:LvM2}
   \end{figure}

Figure \ref{fig:LvM}b contains luminosity \textit{versus} time for the models from \citet{GM2010}'s mass-loss study and an additional early mass-loss model with a smaller initial mass (1.07 $M_{\odot}$). Models include two standard solar models with the GN93, AGS05 abundances, respectively, and early three mass-losing models using the AGS05 abundances and initial masses of 1.07, 1.15 and 1.30 $M_{\odot}$. The luminosity at early times is higher than the present solar luminosity for all early mass-loss models. The early mass-losing models asymptotically approach the standard solar models around 2 Gyrs. Using the luminosity and mass at $t_{1}$ and $t_{2}$ for each of the models in Figure \ref{fig:LvM}b, the constraints on solar luminosity can be compared to the models at these times (Figure \ref{fig:LvM2}). 
   
Figure \ref{fig:LvM2}a compares the luminosity \textit{versus} mass constraints described above to the mass and luminosity at times $t_{1}$ and $t_{2}$ for the standard solar model using the AGS05 abundances and models with initial masses of 1.07 and 1.15 $M_{\odot}$. As expected, the luminosity of the standard solar model at $t_{1}$ is lower than what is required to photodissociate water on Earth (\textit{blue square}). Yet, this luminosity is also too low for there to even be liquid water on the surface of Earth at $t_{1}$ \citep{Kasting_2010}. For the model with an initial mass of 1.07 $M_{\odot}$, luminosity remains low enough to keep liquid water on Earth (\textit{blue diamond}), while water would photodissociate for the model with an initial mass of 1.15 $M_{\odot}$ (\textit{blue triangle}). This rules out all models with a mass greater or equal to 1.15 $M_{\odot}$, with this mass-loss prescription. As expected, the luminosity of the standard solar model at $t_{2}$ would result in frozen and not liquid water on Mars (\textit{red square}), while models with initial masses of 1.07 and 1.15 $M_{\odot}$ have a high enough luminosity at $t_{2}$ to allow for liquid water on Mars. 

Figure \ref{fig:LvM2}b shows luminosity \textit{versus} time for the models from \citet{GM2010}'s mass-loss study and an additional early mass-loss model with a lower initial mass (1.07 $M_{\odot}$), and the constraints derived for each of the models, given the mass at time t. The luminosity trace is greater than the symbol representing the constraint at that $t_{1}$ and mass for models with initial masses 1.15 and 1.30 $M_{\odot}$, meaning water would photodissociate from Earth for these models, while the luminosity trace is less than the symbol representing the constraint at that $t_{1}$ and mass for models with initial masses 1.00 and 1.07 $M_{\odot}$. Similarly, at $t_{2}$ the luminosity trace is less than the symbol representing the constraint at that $t_{2}$ for the standard solar model, meaning that water would be frozen on Mars, while the luminosity trace is greater than the symbol representing the constraint at that $t_{2}$ and mass for models with initial masses 1.07, 1.15, and 1.30 $M_{\odot}$. 

Considering these constraints suggests that the initial mass of the Sun was less than 1.15 $M_{\odot}$ and greater than 1.00 $M_{\odot}$. Models with an initial mass of 1.30 $M_{\odot}$ have been ruled out by the constraints, but they are included in the remainder of the discussion to illustrate the tendency of the solar models with increasing initial mass. The exact mass limits would change based on the mass loss treatment used in the solar evolution model; however the constraints remain fixed given that the orbital eccentricities and angular momentum per unit mass of the planet remain constant with time and changing solar mass. Because mass-loss rates in young solar-like stars remain an open question, with possible rates ranging between 1 x 10$^{-12}$ $M_{\odot}$ yr$^{-1}$ and  1 x 10$^{-9}$ $M_{\odot}$ yr$^{-1}$ \citep{Caffee87,GGB00,Wood02, Wood05a, Wood05b,Cranmer17,Fichtinger17}, additional investigations, like the work of \citet{Minton2007}, examining how the initial mass limit changes with different mass-loss treatments are worthwhile.

\subsection{Impact of Mass Loss on Helioseismic Indicators}
 \begin{figure}    %%%%%%%%%%%%%%%%%% FIGURE 3
                                % includes the two top panels 
   \centerline{
               \includegraphics[width=0.49\textwidth,clip=]{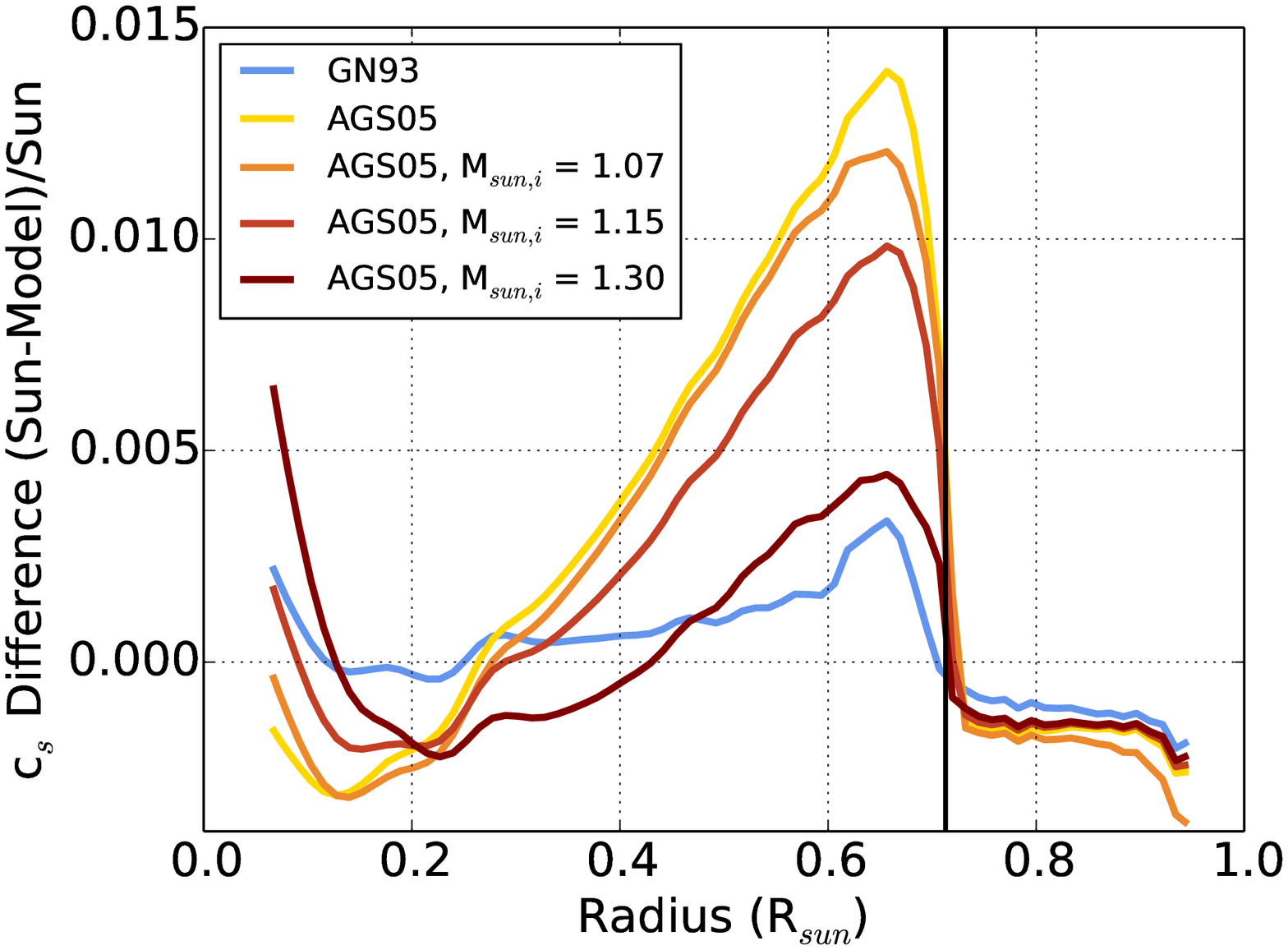}
               \hspace{0.02\textwidth} 
               \includegraphics[width=0.49\textwidth,clip=]{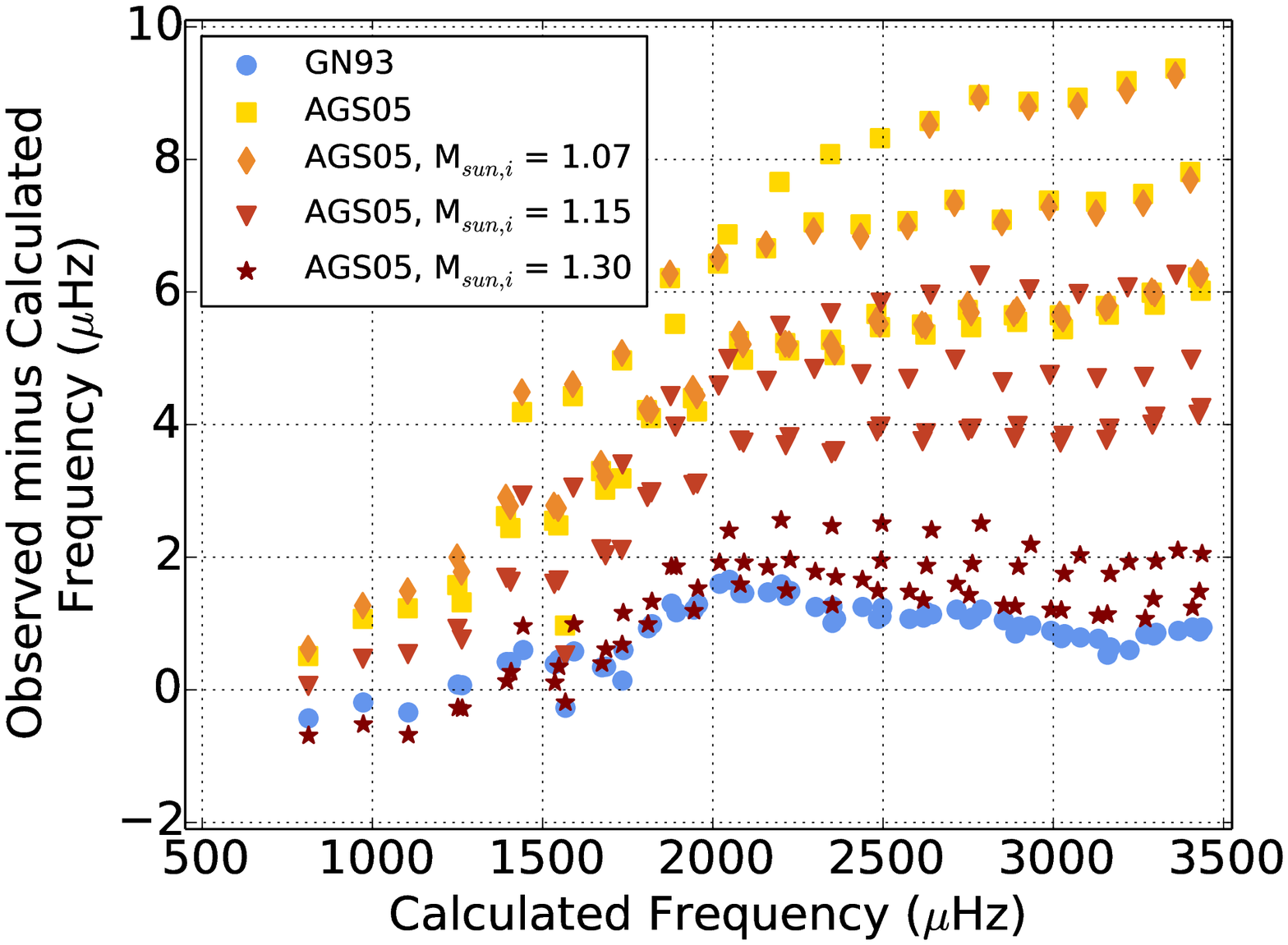}
              }
     \vspace{-0.40\textwidth}   % Shift close to the panel top 
     \centerline{\bf     % Includes the labels (here needs the color 
                                %   package, see beginning of this file)
      \hspace{0.0 \textwidth}  \color{black}{(a)}
      \hspace{0.415\textwidth}  \color{black}{(b)}
         \hfill}
     \vspace{0.38\textwidth}    % Shift back to the panel bottom 
   \centerline{
               \includegraphics[width=0.49\textwidth,clip=]{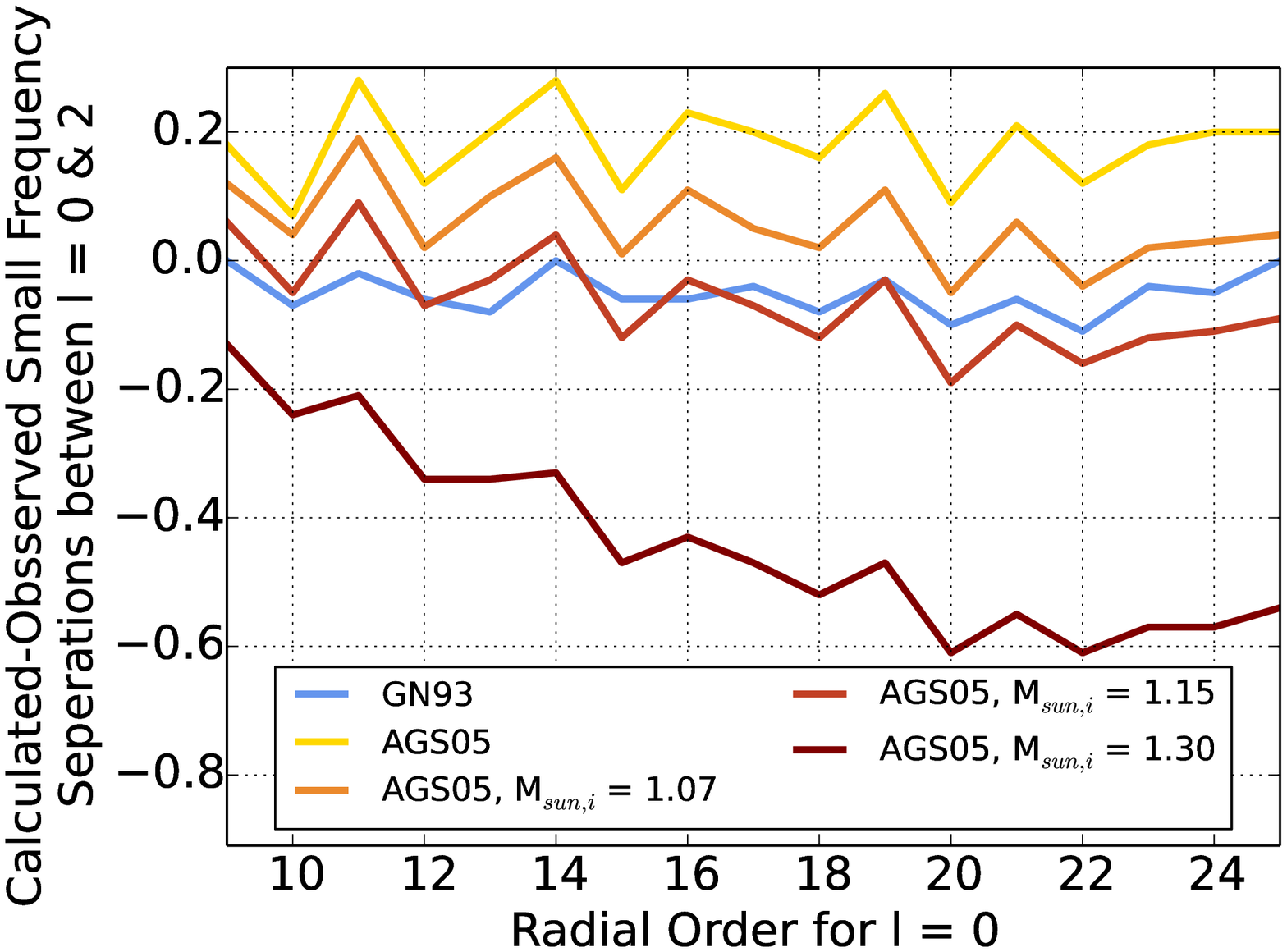}
              }
     \vspace{-0.39\textwidth}   % Shift close to the panel top 
     \centerline{\bf     % Includes the labels (here needs the color 
                                %   package, see beginning of this file)
      \hspace{0.20\textwidth}  \color{black}{(c)}
         \hfill}
     \vspace{0.37\textwidth}    % Shift back to the panel bottom 

 \caption{Comparison between helioseismic data and models. Models include standard solar models with GN93 (\textit{blue}) and AGS05 (\textit{yellow}) abundances and early mass losing models using the AGS05 abundances and initial masses of 1.07, 1.15, and 1.30 $M_{\odot}$ (\textit{orange to maroon}). The GN93, AGS05, mass loss models with initial masses of 1.15 and 1.30 $M_{\odot}$ are from \citet{GM2010}. a) Differences between observed and calculated sound speed for calibrated solar models. The \textit{black vertical line} at $R$ = 0.713 $R_{\odot}$ marks the seismically inferred convection zone base \citep{BA04a}. b) Differences between observed and calculated frequency \textit{versus} calculated frequency. Modes $\ell$=0, 2, 10, and 20 are shown. The calculated frequencies were computed using the \citet{Pesnell} non-adiabatic stellar pulsation code. The data were taken from \citet{Chaplin_2007, ST96,Garcia_2001}. c) Calculated minus observed small separations for modes $\ell$=0 and 2.}
   \label{fig:ml}
   \end{figure}
   
Here we extend the study of helioseismic predictions for mass-losing models conducted by \citet{GM2010} to include a model with a slightly lower initial mass. Figure \ref{fig:ml}a compares the inferred minus calculated sound speed for three mass-losing models to the inferred minus calculated sound speed for standard solar models using GN93 abundances and the AGS05 abundances. The mass-losing models incorporating the AGS05 abundances with initial masses of 1.07 $M_{\odot}$ (ML107), 1.15 $M_{\odot}$ (ML115), and 1.30 $M_{\odot}$ (ML130). The GN93, AGS05, ML115, and ML130 models were taken from \citet{GM2010}. A vertical line marks the seismically inferred convection-zone base radius from \citet{BA04a} at 0.713 $\pm $ 0.001 $R_{\odot}$. As the initial mass of the early mass-loss model increases, agreement between the inferred and calculated sound speeds improves within the radiative zone, below the base of the convection zone but above the core of the Sun. Agreement between the inferred and calculated sound speed is improved in the core of the Sun for the ML107 model, whereas agreement between the inferred and calculated sound speed in the core was decreased for the ML115 and ML130 models.

The observed minus calculated \textit{versus} calculated non-adiabatic frequencies for modes of angular degree $\ell$ = 0, 2, 10, and 20 are shown in Figure \ref{fig:ml}b. These modes propagate into the solar interior below the convection zone. The calculated frequencies were computed using the \citet{Pesnell} non-adiabatic stellar pulsation code. The data were taken from \citet{Chaplin_2007, ST96} and \citet{Garcia_2001}; the \citet{ST96} data were taken using the LOWL instrument \citep{LOWL} and can be found in Table \ref{Tab:ST96}.  As the initial mass of the models increases, the differences between the observed and calculated frequency decreases. This suggests that in the region probed by modes of angular degree $\ell$ = 0, 2, 10, and 20, a higher initial mass improves helioseismic agreement at the present time.

%%%%%%%%%%%%%%%%%%%%%%%%%%%%%%%%%%%%%%%%%%%%%%%%%%%%%%%%%%%%%%%%%%%%%%%%%%%%%
\begin{table}
\caption{Observed Frequencies from LOWL Instrument}
\label{Tab:ST96}
\begin{tabular}{rccrcc}     % define the column alignment
                           % l: left, c: center, r: right
  \hline                   % horizontal line
  \hline 
   $\ell$ = 10\tabnote{Data taken using the LOWL instrument (Tomczyk et al., 1995) that can be found at Schou and Tomczyk (1996)} & &  &  $\ell$ = 20$^{1}$ &  & \\
$ n$ & Frequency & Standard Deviation & $ n $ & Frequency & Standard Deviation  \\
  & ($\mu$Hz)  & ($\mu$Hz) &  &  ($\mu$Hz) & ($\mu$Hz)  \\
\hline
6 & 1443.4270 & 0.0067 & 5 & 1566.5555 & 0.0110  \\
7 & 1592.9194 & 0.0067 & 6 & 1734.1945 & 0.0085 \\
8 & 1737.8435 & 0.0067 & 7 & 1893.9236 & 0.0120 \\
9 & 1880.0386 & 0.0106 & 8 & 2050.0850 & 0.0120 \\
10 & 2021.5305 & 0.0103 & 9 & 2203.1387 & 0.0131 \\
11 &  2162.9443 & 0.0120 & 10 & 2352.4209 & 0.0124  \\
12 & 2302.8081 & 0.0113 & 11 & 2498.9675 & 0.0120  \\
13 & 2440.8206  & 0.0113 & 12 & 2644.8384 & 0.0113  \\
14 & 2578.6360 & 0.0099 & 13 & 2790.6719 & 0.0103 \\
15 & 2716.9546 & 0.0088 & 14 & 2935.9194 & 0.0106  \\
16 & 2855.6741 & 0.0085 & 15 & 3080.4194 & 0.0106 \\
17 & 2994.4666 & 0.0078 & 16 & 3224.3574 & 0.0131  \\
18 & 3132.8979 & 0.0088 & 17 & 3368.0244 & 0.0173 \\
19 & 3271.3110 & 0.0117 &  &   &  \\
20 & 3410.1030 & 0.0163 &  &  &   \\
  \hline
\end{tabular}
\end{table}

%%%%%%%%%%%%%%%%%%%%%%%%%%%%%%%%%%%%%%%%%%%%%%%%%%%%%%%%%%%%%%%%%%%%%%%%%%%

Figure \ref{fig:ml}c contains the small frequency separations of the standard models and the mass-losing models minus the solar-cycle frequency separations from the BiSON group \citep{Chaplin_2007} for the $\ell$ = 0 and 2 modes, which are sensitive to the structure of the core. Differences between calculated and observed small separations for $\ell$ = 0 and 2 modes are decreased for mass-losing models with intermediate initial mass (1.07 - 1.15 $M_{\odot}$), which interestingly is the same mass range relevant to the planetary constraints. 

\subsection{Combined Effects of Mass Loss and Dynamic Screening}
 \begin{figure}    %%%%%%%%%%%%%%%%%% FIGURE 4
                                % includes the two top panels 
   \centerline{
               \includegraphics[width=0.49\textwidth,clip=]{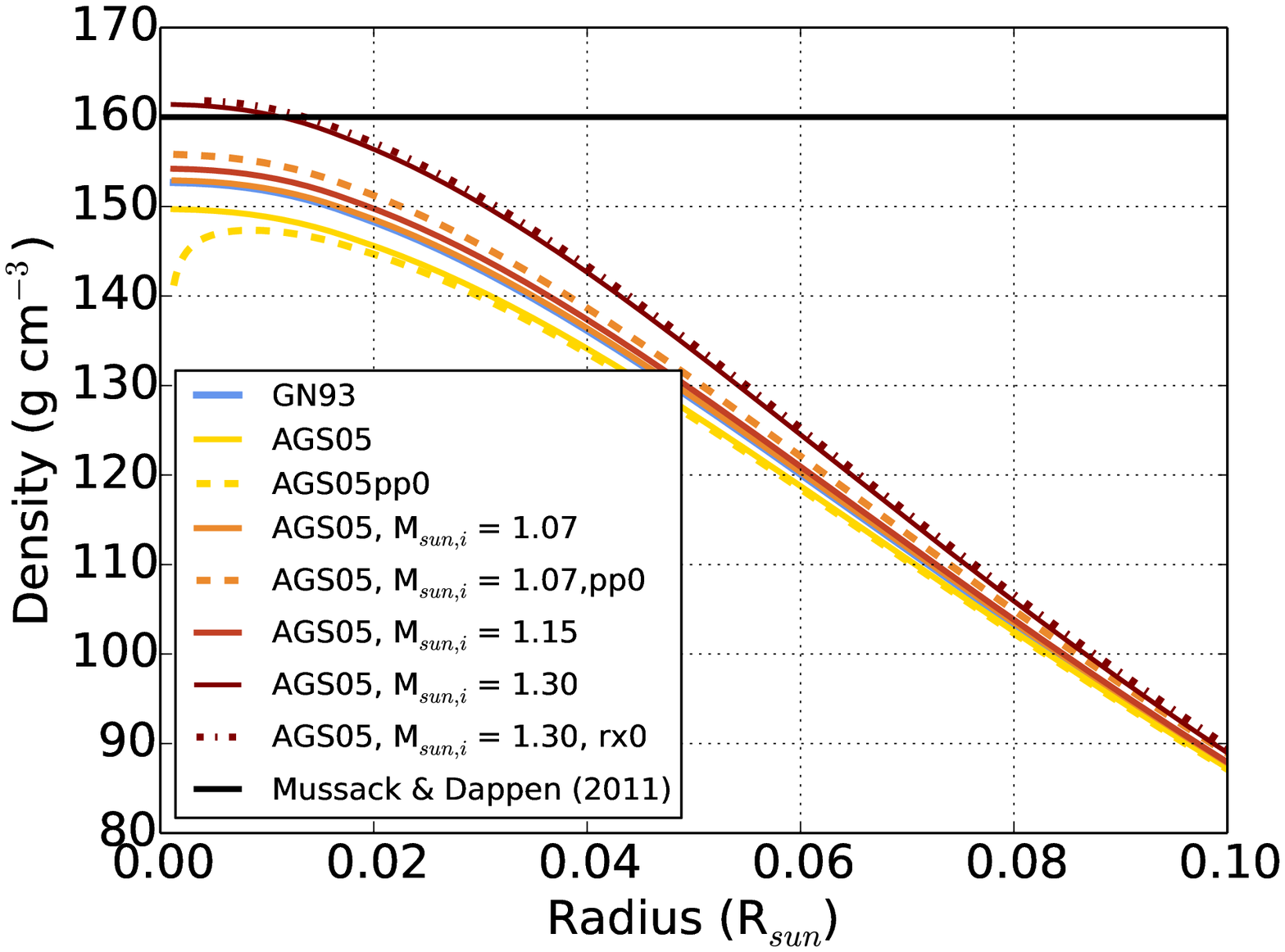}
               \hspace{0.02\textwidth} 
               \includegraphics[width=0.49\textwidth,clip=]{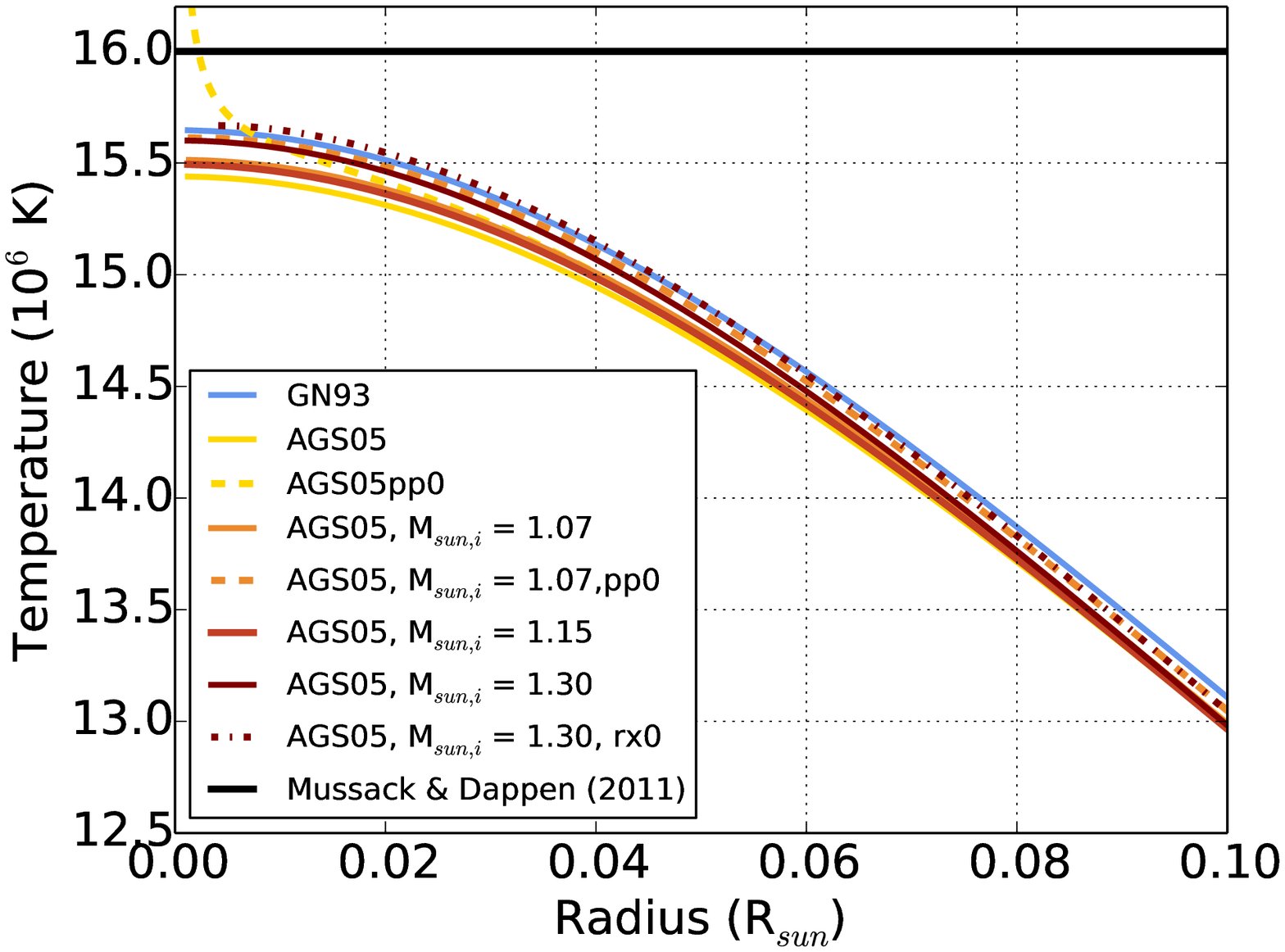}
              }
     \vspace{-0.40\textwidth}   % Shift close to the panel top 
     \centerline{\bf     % Includes the labels (here needs the color 
                                %   package, see beginning of this file)
      \hspace{0.0 \textwidth}  \color{black}{(a)}
      \hspace{0.415\textwidth}  \color{black}{(b)}
         \hfill}
     \vspace{0.38\textwidth}    % Shift back to the panel bottom 
   \centerline{
               \includegraphics[width=0.49\textwidth,clip=]{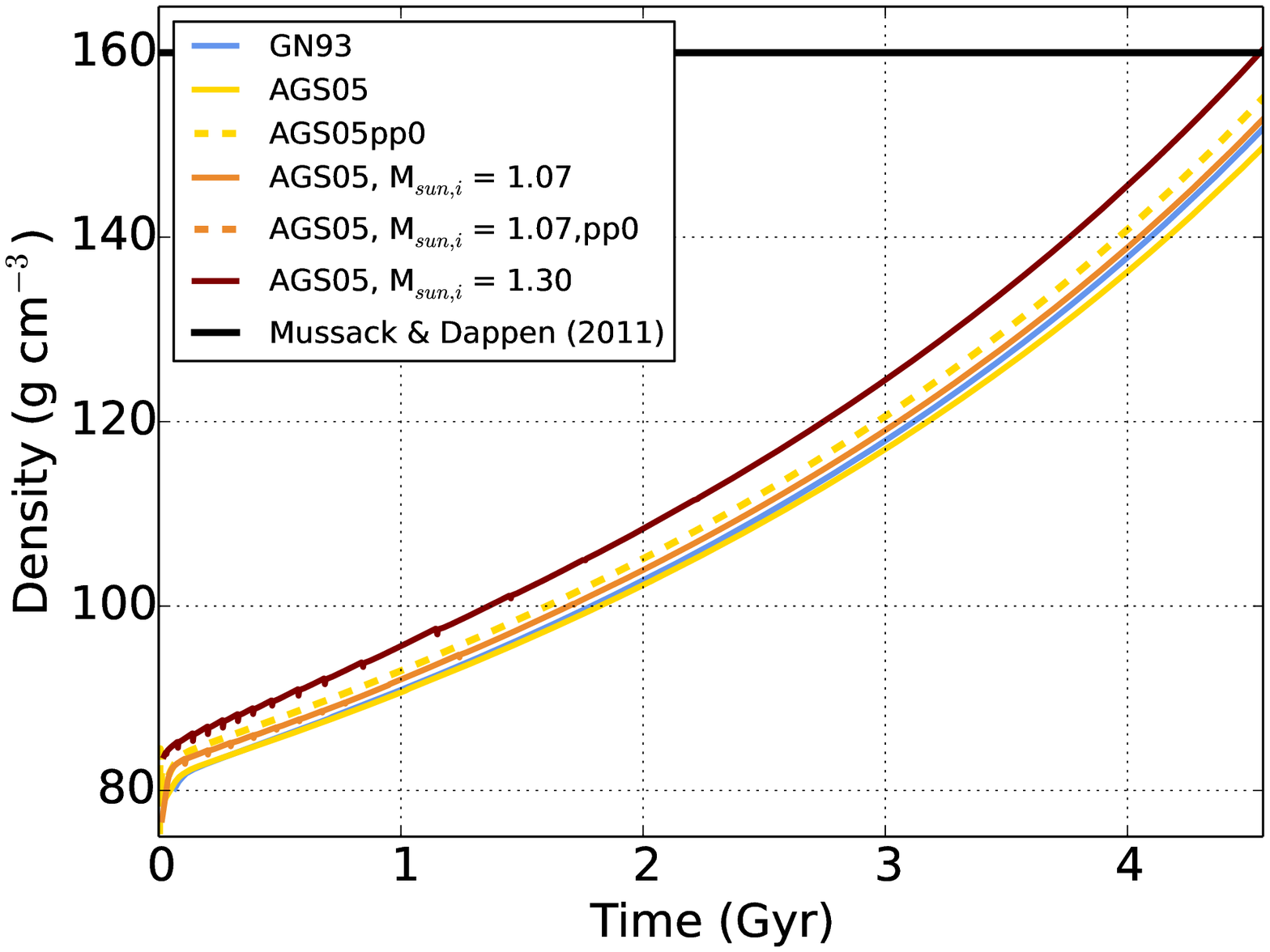}
               \hspace*{0.02\textwidth}
               \includegraphics[width=0.49\textwidth,clip=]{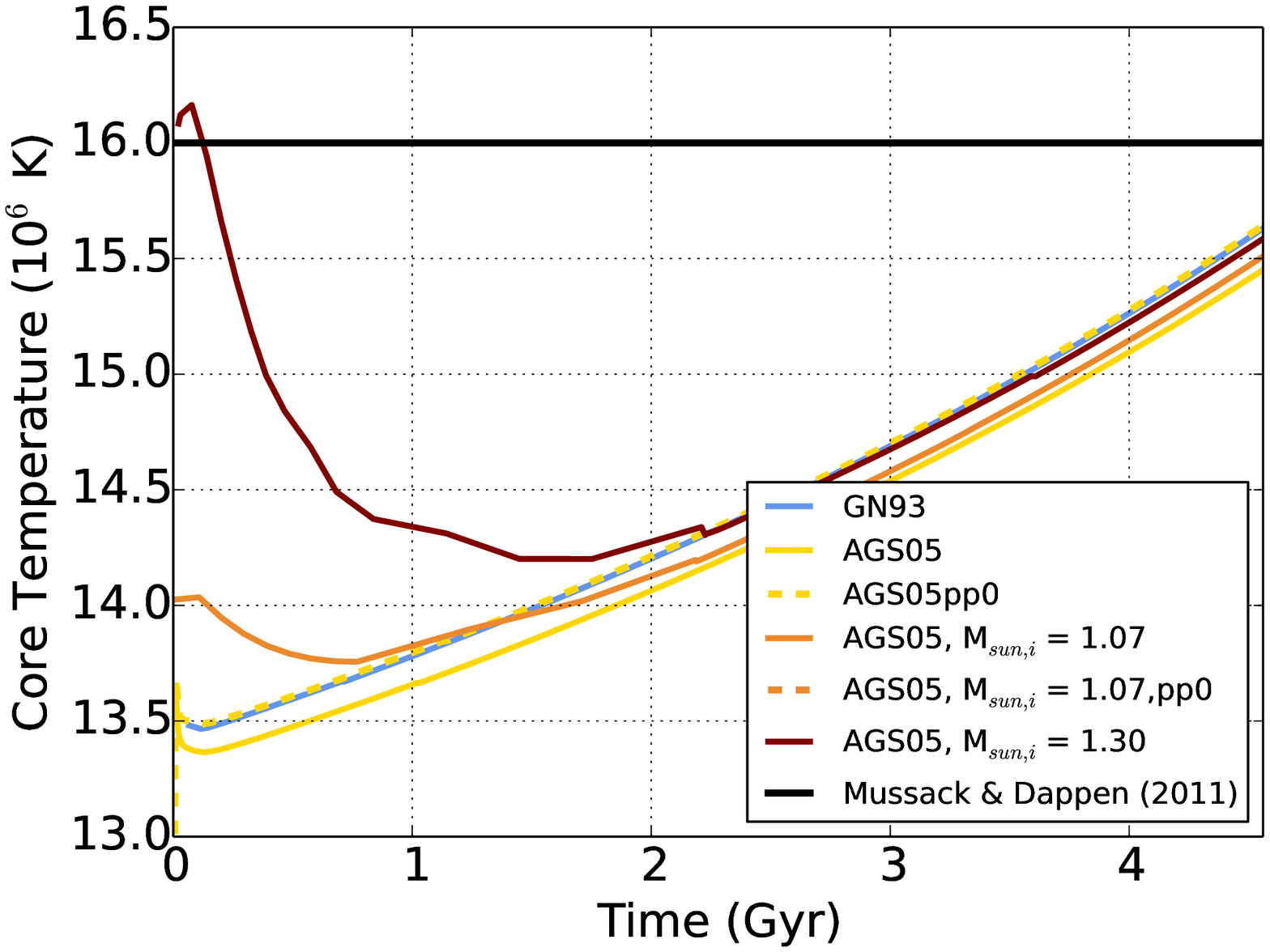}
              }
     \vspace{-0.40\textwidth}   % Shift close to the panel top 
     \centerline{\bf     % Includes the labels (here needs the color 
                                %   package, see beginning of this file)
      \hspace{0.0 \textwidth}  \color{black}{(c)}
      \hspace{0.415\textwidth}  \color{black}{(d)}
         \hfill}
     \vspace{0.37\textwidth}    % Shift back to the panel bottom 

 \caption{(a) Density and (b) temperature \textit{versus} solar radius. Models include: standard solar models with GN93 (\textit{blue}) and AGS05 (\textit{yellow}) abundances, early mass-losing models using the AGS05 abundances and initial masses of 1.07 and 1.30 $M_{\odot}$ (orange and maroon), and early-mass loss models with dynamic screening at the present solar age using the AGS05 abundances and initial masses of 1.07 and 1.30 $M_{\odot}$ (dashed lines of same color). Conditions for electron screening approximation used by \citet{Mussack_2011} are marked with a thick black line.}
   \label{fig:MDcomp}
   \end{figure}

\citet{Mussack_2011} implemented a dynamic electron screening correction to a solar evolution model utilizing AGS05 abundances, and found improved agreement between the inferred and calculated sound speeds in the core of the Sun. As seen in Figure \ref{fig:ml}, early mass loss impacts the sounds speed and oscillation modes in the core of the Sun. Here we will investigate the combined effects of early mass loss and a dynamic electron screening correction. It was expected that agreement between the inferred and the calculated sound speed will improve in both the core and the radiative zone for mass-losing models with dynamic screening corrections. 

We explored different implementations of the dynamic electron screening correction. For the standard solar model using the AGS05 abundances (AGS05pp0, \textit{yellow dashed}) and an early mass-loss model with an initial mass of 1.07 $M_{\odot}$ and the AGS05 abundances (ML107pp0, \textit{orange, dashed}), the dynamic electron screening correction was implemented into the Iben evolution codes by simply removing the \citet{Salpeter_1954} screening enhancement, and treating all p-p reactions as unscreened reactions, because the reaction rate for unscreened and dynamically screened p-p reactions is about the same \citep{Mussack_2011}. To implement an approximate dynamic electron screening correction for the ML130rx0 model (\textit{maroon, dash-dot}), an early mass-loss model with an initial mass of 1.30 $M_{\odot}$ using the AGS05 abundances,  p-p, $^{12}$C+p, and $^{14}$N+p reactions were treated as unscreened reactions by removing the Salpeter screening enhancement for these reactions; the CNO reactions, in particular  $^{12}$C+p and $^{14}$N+p, are important for higher initial-mass models and thus only implemented for this higher initial mass model. While the 1.30 $M_{\odot}$ solar model was ruled out by luminosity constraints, the impact of dynamic screening in higher initial mass stellar models remains as motivation for this implementation. The inclusion of the CNO reactions was done even though the molecular dynamics work was only completed for p-p reactions. The conditions used by \citet{Mao_2009} and \citet{Mussack_2011} are hotter and denser than the conditions in the core of the solar models, for the majority of the evolution (Figure \ref{fig:MDcomp}). Figures \ref{fig:MDcomp}a and \ref{fig:MDcomp}b show the present day conditions for each of the solar models.  Figures \ref{fig:MDcomp}c and \ref{fig:MDcomp}d display the density and temperature, respectively, of the innermost zone of the core as a function of time. The density only approaches the conditions used by \citet{Mao_2009} and \citet{Mussack_2011} at present time; only the model with an initial mass of 1.30 $M_{\odot}$ has a temperature near the conditions used by \citet{Mao_2009} and \citet{Mussack_2011} at a time other than present.

  \begin{figure}    %%%%%%%%%%%%%%%%%% FIGURE 5
                                % includes the two top panels 
   \centerline{
               \includegraphics[width=0.49\textwidth,clip=]{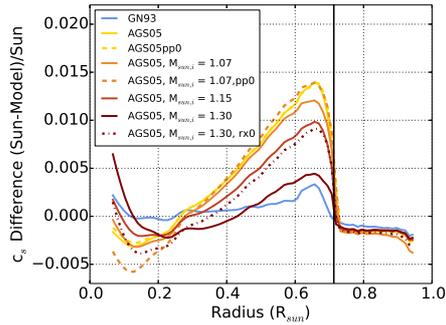}
              }

\caption{Relative difference between inferred and calculated sound speeds. Models include: standard solar models with GN93 and AGS05 abundances, a model with dynamic screening model correction for p-p reactions using the AGS05 abundances, early mass-losing models using the AGS05 abundances and initial masses of 1.07 and 1.30 $M_{\odot}$, and early-mass loss models with dynamic screening correction at the present solar age using the AGS05 abundances and initial masses of 1.07 and 1.30 $M_{\odot}$. For the model with an initial mass of 1.07 $M_{\odot}$, the screening correction is applied to p-p reactions. For the model with an initial mass of 1.30 $M_{\odot}$, the screening correction is applied to p-p reactions as well as $^{12}$C+P and $^{14}$N+p reactions.
        }
   \label{fig:cs_ML107vspp0}
   \end{figure}

Figure \ref{fig:cs_ML107vspp0} shows the inferred minus calculated sound speed for two standard solar models (\textit{solid line}), three mass-losing models (\textit{solid line}), and two mass-losing models with screening corrections (\textit{dashed line}). The GN93, AGS05, ML107, and ML130 models are the same models shown in Figure \ref{fig:ml}. All models except the GN93 (\textit{blue}) standard solar model incorporate the AGS05 abundances (\textit{yellow to maroon}). The dynamic screening correction approximation for only p-p reactions for the AGSpp0 and ML107pp0 models (\textit{dashed}). The ML130rx0 model incorporates the dynamic screening correction approximation for the p-p reactions, as well as the $^{12}$C+p and $^{14}$N+p reactions of the CNO cycle (\textit{dash-dot}). While the incorporation of the dynamic screening correction slightly improves agreement between the inferred and calculated sound speeds for the standard model (\textit{yellow, solid vs. dashed}), the incorporation of the same approximation worsens agreement in the core and below the convection zone of Sun for the ML107pp0 model (\textit{orange, solid vs. dashed}) (Figure \ref{fig:cs_ML107vspp0}). The implementation of the approximate dynamic screening correction for the ML130rx0, improves sound-speed agreement in the core of the Sun as compared to the ML130 model, while it diminishes agreement with helioseismic data for the radiative zone above the core as compared to the ML130 model (Figure \ref{fig:cs_ML107vspp0}). The dynamic screening correction has a greater impact on the sound-speed agreement of the ML130rx0 model than the ML107pp0 model (Figure \ref{fig:cs_ML107vspp0}), which points to the need for a molecular dynamics study for CNO reactions. 

\begin{figure}    %%%%%%%%%%%%%%%%%% FIGURE 6
                                % includes the two top panels 
   \centerline{
               \includegraphics[width=0.49\textwidth,clip=]{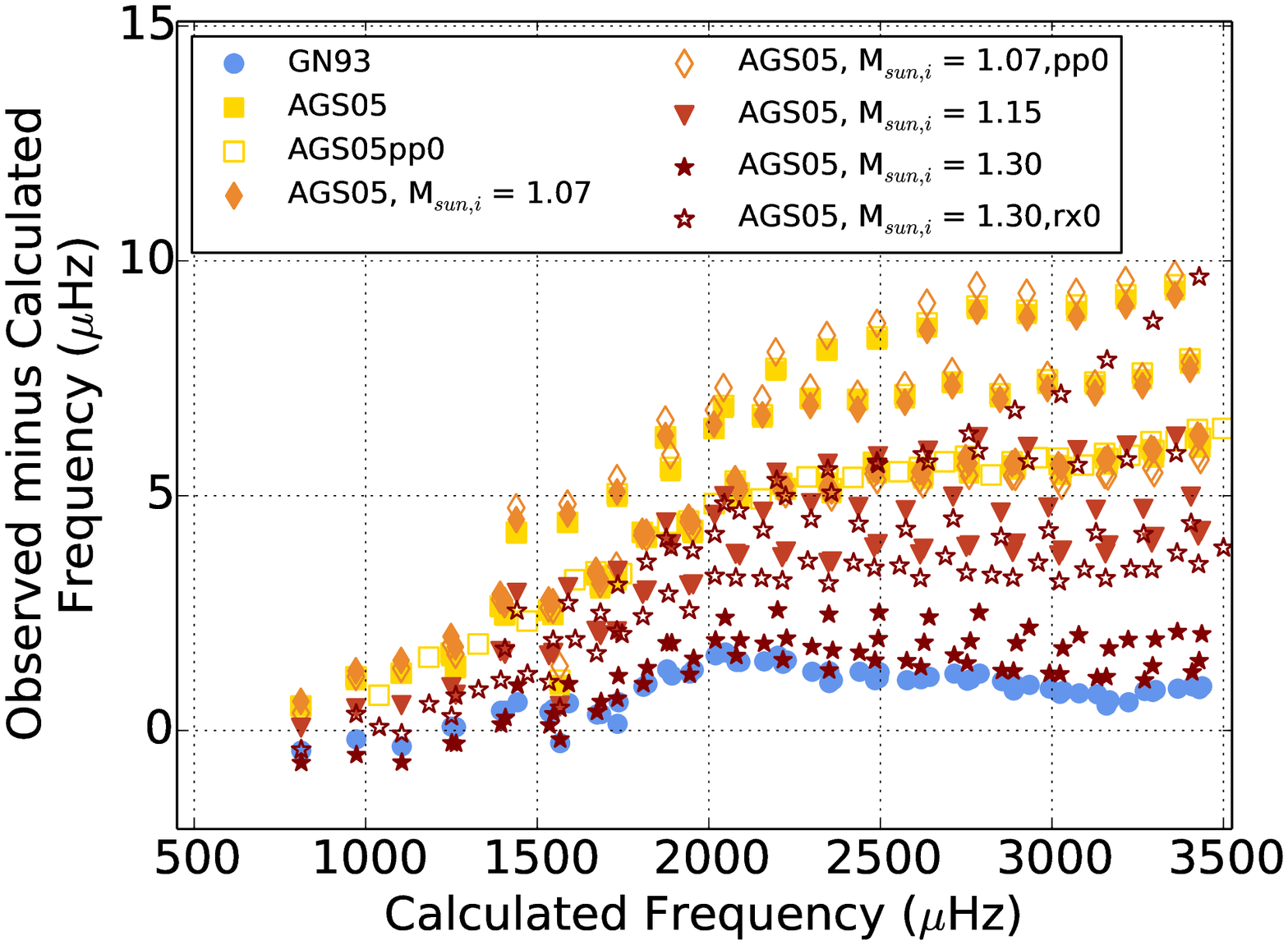}
                \hspace*{0.02\textwidth}
               \includegraphics[width=0.49\textwidth,clip=]{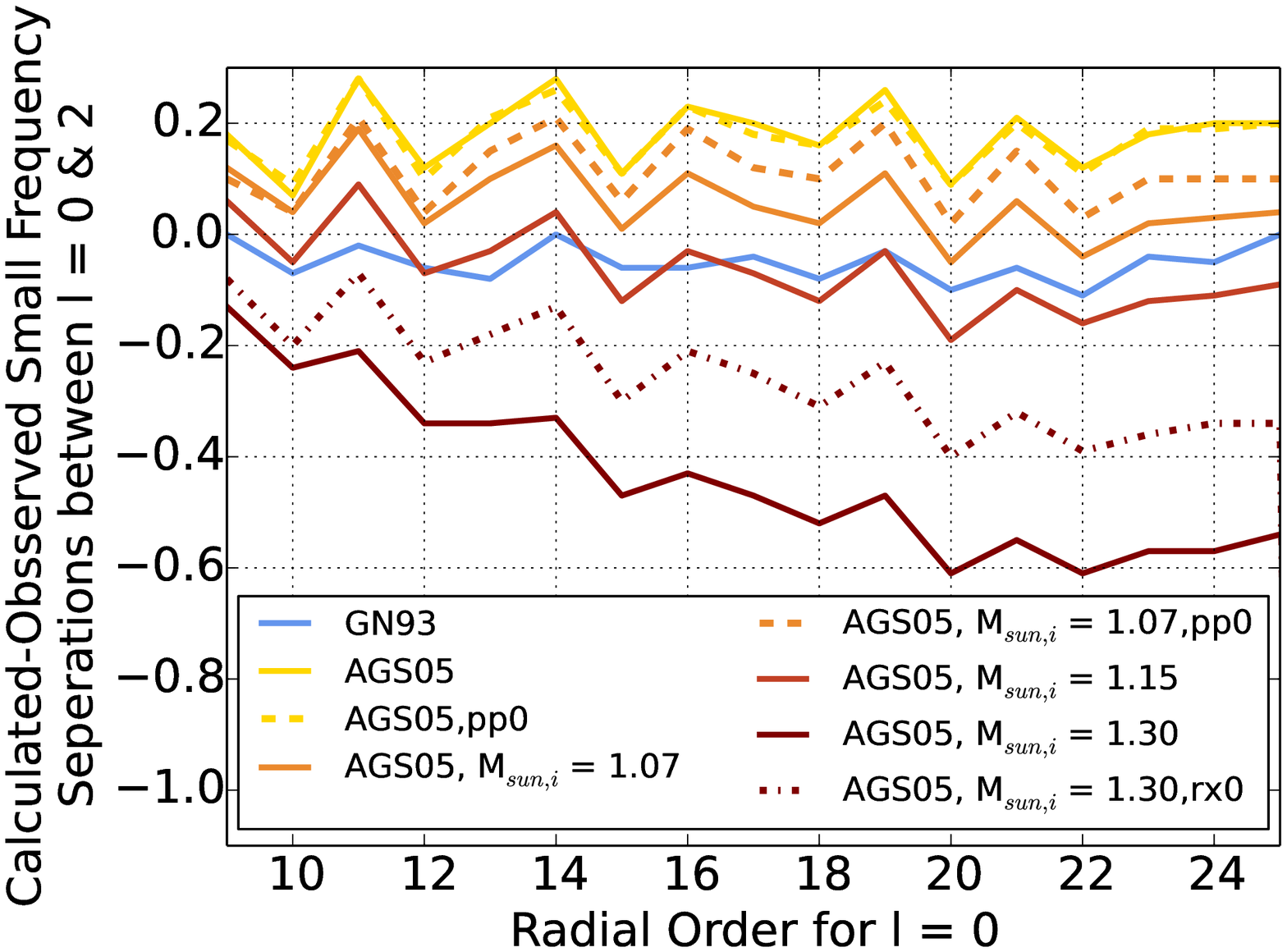}
              }
     \vspace{-0.40\textwidth}   % Shift close to the panel top 
     \centerline{\bf     % Includes the labels (here needs the color 
                                %   package, see beginning of this file)
      \hspace{0.0 \textwidth}  \color{black}{(a)}
      \hspace{0.415\textwidth}  \color{black}{(b)}
         \hfill}
     \vspace{0.37\textwidth}    % Shift back to the panel bottom 

 \caption{(a) Differences between observed and calculated frequency \textit{versus} calculated frequency for modes $\ell$=0, 2, 10, and 20 modes. Models include: standard solar models with GN93 and AGS05 abundances, a dynamic screening model with correction for pp-reactions using the AGS05 abundances, early mass-losing models using the AGS05 abundances and initial masses of 1.07 and1.30 $M_{\odot}$, and early-mass loss models with dynamic screening at the present solar age using the AGS05 abundances and initial masses of 1.07 and 1.30 $M_{\odot}$. The calculated frequencies were computed using the \citet{Pesnell} non-adiabatic stellar pulsation code. The data are taken from \citet{Chaplin_2007}, \citet{ST96}, and \citet{Garcia_2001}. (b) Difference between calculated and observed small separations for modes $\ell$=0 and 2. Models from the \citet{GM2010} models using the GN93 abundances and the AGS05 abundances and mass loss with initial mass of 1.07 $M_{\odot}$ and 1.30 $M_{\odot}$ are compared to the corresponding models with dynamic screening.}
   \label{fig:ominc_ML107vspp0}
   \end{figure}

The observed minus calculated \textit{versus} calculated non-adiabatic frequencies for modes of angular degree $\ell$ = 0, 2, 10, and 20, which propagate into the solar interior below the convection zone, are shown in Figure \ref{fig:ominc_ML107vspp0}a. The dynamic screening correction for the Ml107pp0 model slightly decreases agreement in the radiative zone as seen by the slight increase in observed minus calculated frequencies (Figure \ref{fig:ominc_ML107vspp0}a, orange). The dynamic screening correction for the ML130rx0 model decreases the agreement in the radiative zone compared to the ML130 model, as seen by increase in observed minus calculated frequencies (Figure \ref{fig:ominc_ML107vspp0}a, maroon). The dynamic screening correction has a greater effect on the observed minus calculated non-adiabatic frequencies for the ML130rx0 model than the ML107pp0 model (Figure \ref{fig:ominc_ML107vspp0}a \textit{maroon and orange}, respectively).

Figure \ref{fig:ominc_ML107vspp0}b shows the small frequency separations of the models discussed above minus the observed small frequency separations from the BiSON group \citep{Chaplin_2007} for the $\ell$ = 0 and 2 modes, which are sensitive to the structure of the core. While the decrease in sound-speed agreement in the core of the Sun for the ML107pp0 model compared to the ML107 model can be seen in the small frequency separations, which are sensitive to the structure of the core, an improvement in sound-speed agreement in the core of the Sun for the ML130rx0 compared to the ML130 model can be seen in the small frequency separations (Figure \ref{fig:ominc_ML107vspp0}b orange and maroon, respectively). The dynamic screening correction has a greater effect on small frequency separations for the $\ell$ = 0 and 2 modes for the ML130rx0 model than the ML107pp0 model (Figure \ref{fig:ominc_ML107vspp0}b \textit{maroon and orange}).

Although the implementations of the dynamic screening correction approximation are different during the mass-loss phase of the ML107pp0 and ML130rx0 models, due to the fact that CNO-cycle is important for the 1.30 $M_{\odot}$ model until enough mass is lost, the way that this correction affects agreement with helioseismic data is worthy of further investigation, especially in light of the recent observation of gravity modes \citep{Fossat17} which can be used to probe this region. Additionally, there is a need to extend the molecular dynamics dynamic electron screening study to other relevant pressures, temperature, and reactions.

%%%%%%%%%%%%%%%%%%%%%%%%%%%%%%%%%%%%%%%%%%%%%%%%%%%%%%%%%%%%%%%%%%%%%%%%%%%
%% Conclusions
\section{Conclusion}
\label{sect: conclude}
Solar irradiance required to allow liquid water on early Earth and Mars was analyzed in the context of a more massive and luminous early Sun, where the initial solar mass (M$_{\odot,i}$) was determined to be 1.0 $<$ $M_{\odot,i}$ $<$ 1.15 $M_{\odot}$ for an mass-loss model with an exponentially decaying mass-loss rate. Additionally, models with an initial mass between 1.07 - 1.15 $M_{\odot}$ corresponds to better helioseimic agreement in both the core and the radiative zone compared to the standard model. It is fair to conclude that models with an initial mass of 1.07 $M_{\odot}$ and an exponential mass loss rate with an e-folding time of 1/10 of solar age fit both the constraints for early Earth and Mars; however it is possible that models with a somewhat higher or lower M$_{\odot,i}$ are not ruled out by these constraints. The exact mass value will change with different abundances \citep[\textit{e.g.},][]{AGSS09} or with different mass loss laws; the effects of which will be studied in future work. 

In this paper, we approximate dynamic electron screening corrections by removing the Salpeter screening for specific reactions, particularly p-p, as well as  $^{12}$C+p and $^{14}$N+p reactions for higher initial mass stars. The small separations in the solar core are improved for both the intermediate mass and higher initial mass models with specified dynamic screening corrections implementations, 1.07 and 1.30 $M_{\odot}$, respectively. Dynamic electron screening has significant effects and should be taken into account, but more molecular dynamics simulations are needed to derive and implement exact corrections. 

Early mass loss is promising avenue to help explain the presence of liquid water on early Mars and Earth, and also mitigate somewhat the solar abundance problem. But more work needs to be done to explore the consequences for solar Li depletion and solar core rotation \citep[like that of][]{Piau2002, STC10} and to investigate early main sequence mass-loss rates for solar analogs. Recent observations of g modes by \citet{Fossat17} may provide key constraints on the evolution and structure of the solar core. Solar g-modes may provide a way to test and help constrain the initial solar mass and mass-loss history, and the dynamic screening implementations.

%%%%%%%%%%%%%%%%%%%%%%%%%%%%%%%%%%%%%%%%%%%%%%%%%%%%%%%%%%%%%%%%%%%%%%%%%%%
%% Acknowledgments
%
\begin{acks}
The authors thank Sylvaine Turck-Chi\`{e}ze for helpful discussions. The authors would also like to thank Jesper Schou and Steven Tomczyk for providing their data and their permission to list it in this publication. Los Alamos National Laboratory is operated by Los Alamos National Security, LLC for the National Nuclear Security Administration of the U.S. Department of Energy under Contract No. DE-AC52-06NA2539. S.R.W. would like to acknowledge Simon Bolding for his helpful discussions on python scripts.

\textbf{Disclosure of Potential Conflicts of Interest} 
The authors declare that they have no conflicts of interest.

\end{acks}

%%%%%%%%%%%%%%%%%%%%%%%%%%%%%%%%%%%%%%%%%%%%%%%%%%%%%%%%%%%%%%%%%%%%%%%%%%%
%% References
%
\bibliographystyle{spr-mp-sola}
\bibliography{planetCon}

\end{article} 
\end{document}